\documentclass[12pt]{article}

\usepackage{draft} 
\usepackage{hyperref}
\usepackage{graphicx}
\usepackage{cite}
\usepackage{mciteplus}
\usepackage{skak}
\usepackage{empheq}
\usepackage{tikz}
\usepackage{booktabs}
\usepackage{siunitx}
\usepackage{bbm}
\usepackage{makecell}
\usepackage{float}
\usepackage{dsfont}

\usepackage{bm}
\usepackage{braket}

\DeclareFontFamily{OT1}{pzc}{}
\DeclareFontShape{OT1}{pzc}{m}{it}{<-> s * [1.10] pzcmi7t}{}
\DeclareMathAlphabet{\mathpzc}{OT1}{pzc}{m}{it}

\def\be#1\ee{\begin{align}#1\end{align}}

\makeatletter
\newcommand{\bdryno}{\mathpalette\bdry@no\relax}
\newcommand{\bdry@no}[2]{%
  \mspace{1mu}%
  \vbox{%
    \hbox{$\m@th#1\scriptstyle{\ast}$}
    \nointerlineskip
    \kern.25ex
    \hbox{$\m@th#1\scriptstyle{\ast}$}
    \kern-.06ex
  }%
  \mspace{1mu}%
}
\makeatother

\usetikzlibrary{snakes}
\usetikzlibrary{arrows}
\usetikzlibrary{decorations.pathmorphing}
\usetikzlibrary{ decorations.markings}
\tikzset{snake it/.style={decorate, decoration=snake}}
\usetikzlibrary{shapes.misc}
\tikzset{cross/.style={cross out, draw=black, minimum size=2*(#1-\pgflinewidth), inner sep=0pt, outer sep=0pt},
cross/.default={1pt}}
\usetikzlibrary{shapes.geometric}

\definecolor{bleudefrance}{rgb}{0.19, 0.55, 0.91}
\definecolor{candyapplered}{rgb}{1.0, 0.03, 0.0}
\definecolor{myblue}{rgb}{0.0,0.635,1}

\newcommand{\ups}{\Upsilon_{1}}

\newcommand{\cF}{\mathcal{F}}

\newcommand{\Chat}{\widehat{C}}
\newcommand{\Vhat}{\widehat{V}}
\newcommand{\Phat}{\widehat{P}}
\newcommand{\hhat}{\widehat{h}}

\newcommand{\cH}{\mathcal{H}}

\tikzset{
  pics/cylnnrb0/.style n args={2}{
    code = { %
        \filldraw[color=red,fill=black!30, thick] circle (0.35);
        \filldraw[color=blue, fill=white, thick]  circle (0.17);
        \node[color=black] at (0,0.51) {\scriptsize ${#1}$};
        \node[color=black] at (0,0) {\scriptsize ${#2}$};

    }
  }
}

\tikzset{
  pics/cylnnrb/.style n args={2}{
    code = { %
        \filldraw[color=red, densely dashed,fill=black!30, thick] circle (0.35);
        \filldraw[color=blue, densely dashed, fill=white, thick]  circle (0.17);
        \node[color=black] at (0,0.51) {\scriptsize ${#1}$};
        \node[color=black] at (0,0) {\scriptsize ${#2}$};

    }
  }
}
\tikzset{
  pics/cylnnrr0/.style n args={2}{
    code = { %
        \filldraw[color=red,fill=black!30, thick] circle (0.35);
        \filldraw[color=red, fill=white, thick]  circle (0.17);
        \node[color=black] at (0,0.51) {\scriptsize ${#1}$};
        \node[color=black] at (0,0) {\scriptsize ${#2}$};

    }
  }
}
\tikzset{
  pics/cylnnrr/.style n args={2}{
    code = { %
        \filldraw[color=red, densely dashed,fill=black!30, thick] circle (0.35);
        \filldraw[color=red, densely dashed, fill=white, thick]  circle (0.17);
        \node[color=black] at (0,0.51) {\scriptsize ${#1}$};
        \node[color=black] at (0,0) {\scriptsize ${#2}$};

    }
  }
}
\tikzset{
  pics/cylnnrr1/.style n args={2}{
    code = { %
        \filldraw[color=red,fill=black!30, thick] circle (0.35);
        \filldraw[color=red, fill=white, thick]  circle (0.17);
        \node[color=black] at (0,0.51) {\scriptsize ${#1}$};
        \node[color=black] at (0,0) {\scriptsize ${#2}$};
        \node[cross=3pt, very thick] at (-0.25,0) {};
    }
  }
}
\tikzset{
  pics/cylnnbb0/.style n args={2}{
    code = { %
        \filldraw[color=blue,fill=black!30, thick] circle (0.35);
        \filldraw[color=blue, fill=white, thick]  circle (0.17);
        \node[color=black] at (0,0.51) {\scriptsize ${#1}$};
        \node[color=black] at (0,0) {\scriptsize ${#2}$};

    }
  }
}
\tikzset{
  pics/cylnnbb/.style n args={2}{
    code = { %
        \filldraw[color=blue, densely dashed,fill=black!30, thick] circle (0.35);
        \filldraw[color=blue, densely dashed, fill=white, thick]  circle (0.17);
        \node[color=black] at (0,0.51) {\scriptsize ${#1}$};
        \node[color=black] at (0,0) {\scriptsize ${#2}$};

    }
  }
}
\tikzset{
  pics/cylnnbb1/.style n args={2}{
    code = { %
        \filldraw[color=blue,fill=black!30, thick] circle (0.35);
        \filldraw[color=blue, fill=white, thick]  circle (0.17);
        \node[color=black] at (0,0.51) {\scriptsize ${#1}$};
        \node[color=black] at (0,0) {\scriptsize ${#2}$};
        \node[cross=3pt, very thick] at (-0.25,0) {};
    }
  }
}

\tikzset{
  pics/disk1/.style n args={1}{
    code = { %
        \filldraw[color=black, fill=black!30, thick] circle (0.3);
        \node[color=black] at (0,0.5) {\scriptsize ${#1}$};
        \draw node[cross=3pt, very thick] {};
    }
  }
}   
\tikzset{
  pics/disk1r/.style n args={1}{
    code = { %
        \filldraw[color=red, fill=black!30, thick] circle (0.3);
        \node[color=black] at (0,0.5) {\scriptsize ${#1}$};
        \draw node[cross=3pt, very thick] {};
    }
  }
}
\tikzset{
  pics/disk2r/.style n args={1}{
    code = { %
        \filldraw[color=red, fill=black!30, thick] circle (0.3);
        \node[color=black] at (0,0.5) {\scriptsize ${#1}$};
        \node[cross=3pt, very thick] at (0.125,0) {};
        \node[cross=3pt, very thick] at (-0.125,0) {};
    }
  }
}
\tikzset{
  pics/disker/.style n args={1}{
    code = { %
        \filldraw[color=red, fill=black!30, thick] circle (0.3);
        \node[color=black] at (0,0.5) {\scriptsize ${#1}$};
    }
  }
}
\tikzset{
  pics/disk1b/.style n args={1}{
    code = { %
        \filldraw[color=blue, fill=black!30, thick] circle (0.3);
        \node[color=black] at (0,0.5) {\scriptsize ${#1}$};
        \draw node[cross=3pt, very thick] {};
    }
  }
}   
\tikzset{
  pics/disk2b/.style n args={1}{
    code = { %
        \filldraw[color=blue, fill=black!30, thick] circle (0.3);
        \node[color=black] at (0,0.5) {\scriptsize ${#1}$};
        \node[cross=3pt, very thick] at (0.125,0) {};
        \node[cross=3pt, very thick] at (-0.125,0) {};
    }
  }
} 
\tikzset{
  pics/diskeb/.style n args={1}{
    code = { %
        \filldraw[color=blue, fill=black!30, thick] circle (0.3);
        \node[color=black] at (0,0.5) {\scriptsize ${#1}$};
    }
  }
}

\begin{document}

\unitlength = .8mm

\begin{titlepage}

\begin{center}

\hfill \\
\hfill \\
\vskip 1cm

\title{The Torus One-Point Diagram in\\Two-Dimensional String Cosmology}

\author{Victor A. Rodriguez}

\address{
Joseph Henry Laboratories, Princeton University, \\ Princeton, NJ 08544, USA 
}

\email{vrodriguez@princeton.edu}

\end{center}

\abstract{

We calculate numerically the torus one-point string diagram in the two-dimensional string cosmology background by decomposing the one-point functions in $c=1$ and $c=25$ Liouville CFT into torus one-point Virasoro conformal blocks and integrating over the fundamental domain of the torus moduli space. We find a remarkably simple result as a function of the outgoing closed string energy. This torus one-point diagram is expected to contribute to the one-point cosmological wavefunction at order $g_s$, and to the four-point cosmological wavefunction at order $g_s^2$ through the disconnected product of the torus one-point diagram and the sphere three-point diagram.

}

\vfill

\end{titlepage}

\eject

\begingroup
\hypersetup{linkcolor=black}

\tableofcontents

\endgroup

\pagebreak


\section{Introduction} 
\label{sec:intro}


In a recent paper \cite{Rodriguez:2023kkl} a two-dimensional string theory background with explicit time dependence was studied using string perturbation theory. 
The worldsheet conformal field theory (CFT) for this two-dimensional string cosmology is
\begin{equation}
\parbox{.2\textwidth}{\centering $c=1$ \\ Liouville CFT} \oplus \parbox{.2\textwidth}{\centering $c=25$ \\ Liouville CFT} \oplus \parbox{.2\textwidth}{\centering $b,c$ ghosts,} \vphantom{\sum_.^{.}}
\label{eq:wscft}
\end{equation}
where the $c=1$ Liouville CFT sector, suitably continued to Lorentzian signature, is interpreted as the time coordinate of the two-dimensional target spacetime, and the $c=25$ Liouville CFT sector is interpreted as the spatial coordinate of the spacetime. 
At the level of the worldsheet CFT, the continuation to Lorentzian signature is achieved through the continuation of the Virasoro conformal weights of $c=1$ Liouville CFT \emph{external} operators --- that is, of vertex operators that are used to define the on-shell states of the full string theory (\ref{eq:wscft}) --- to imaginary values in order for them to be interpreted as a Lorentzian energy \cite{Bautista:2019jau,Rodriguez:2023kkl}. 
Due to special properties of $c=1$ Liouville CFT, in particular the analytic structure of its three-point function coefficient as well as that of Virasoro conformal blocks at $c=1$, this continuation of the external weights turns out to be trivial \cite{Bautista:2019jau}. 

The spacetime interpretation of the worldsheet string theory (\ref{eq:wscft}), and in particular of the time-dependence of this string theory model, is most apparent in the semiclassical action of the $c=1$ Liouville sector, after continuation to a Lorentzian target spacetime:
\ie
S_{c=1\text{ Liouv.}}[\chi^0] = \frac{1}{4\pi} \int d^2\sigma	\sqrt{g} \left( -g^{mn}\partial_m\chi^0\partial_n\chi^0 + 4\pi\mu e^{-2\chi^0} \right).
\label{eq:timeLiouv}
\fe
The exponential potential term $e^{-2\chi_0}$ provides the explicit time-dependence of the closed string theory background (\ref{eq:wscft}). 
In the infinite past $\chi_0\to -\infty$, the potential is exponentially large which signals that perturbative excitations of closed strings are exponentially heavy. 
On the other hand, in the infinite future $\chi_0\to\infty$ the time-dependent potential vanishes and we recover the usual time-independent two-dimensional string theory background (also commonly referred to as ``$c=1$ string theory" \cite{Klebanov:1991qa,Ginsparg:1993is,Jevicki:1993qn,Polchinski:1994mb,Martinec:2004td,Nakayama:2004vk,Balthazar:2017mxh,Balthazar:2019rnh}). 
In the infinite future, on-shell single particle string states are represented by the vertex operators
\ie
\cV_\omega \,=\, g_s\, \Vhat_{-i\omega} \, V_{P=\omega},
\label{eq:vertexop}
\fe
where $\Vhat_{P_{\rm ext}}$ denotes a Virasoro primary of weight $\hhat=P_{\rm ext}^2$ in the $c=1$ Liouville sector of (\ref{eq:wscft}), and $V_{P}$ denotes a Virasoro primary of weight $h=1+P^2$ in the $c=25$ Liouville sector of (\ref{eq:wscft}).\footnote{See \cite{Rodriguez:2023kkl} for a more comprehensive review of $c=1$ and $c=25$ Liouville CFTs, whose notation we will follow in this paper.} 

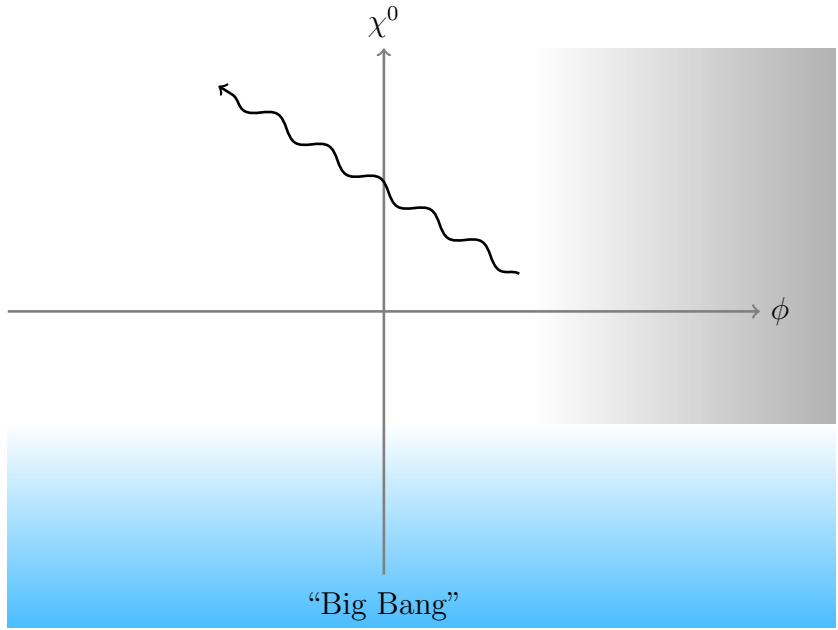
\begin{figure}[h!]
\centering
\begin{tikzpicture}
\shade[left color = black!0,right color = black!30] (2,-3.5) rectangle (6,3.5);
\shade[top color = myblue!0, bottom color = myblue!70] (-5,-4.2) rectangle (6,-1.5);
\draw[->,color=gray, line width=1pt] (-5,0) -- (5,0);
\node[right] at (5,0) {$\phi$};
\draw[->,color=gray, line width=1pt] (0,-3.5) -- (0,3.5);
\node[above] at (0,3.5) {$\chi^0$};
\draw[->, color=black, line width=1pt, decorate, decoration={snake, segment length=8mm, amplitude=1mm}] (1.8,0.5) -- (-2.2,3);
\node[above] at (0,-4.25) {``Big Bang"};
\end{tikzpicture}
\caption{Spacetime interpretation of the two-dimensional string cosmology background (\ref{eq:wscft}). The region shaded in gray represents the spacelike Liouville potential wall, whereas the region shaded in blue represents the time-dependent Liouville potential $e^{-2\chi^0}$. Cosmological wavefunctions in string perturbation theory compute the overlap of an initial state in the infinite past, denoted by ``Big Bang", and an out-state in the Hilbert space of perturbative string states. 
Represented in the picture, the cosmological one-point function $\Psi(\omega)$ is a measure of the production of a single particle state with energy $\omega$ in the far future in this time-dependent background.  
}
\label{fig:spacetime}
\end{figure}

In this time-dependent string theory background, string perturbation theory does not compute an S-matrix element of perturbative in- and out- string states. 
Instead, \cite{Rodriguez:2023kkl} interpreted string perturbation theory as computing an overlap between the initial state of the two-dimensional universe (or spacetime) in the infinite past with a multi-particle state in the Hilbert space of perturbative string states in the infinite future. 
In this way, string perturbation theory probes particular components (in the basis of perturbative strings in the far future) of the initial wavefunction of the universe. 
For short, we will refer to this kind of observable as a cosmological wavefunction. 
See Figure \ref{fig:spacetime} for a pictorial representation of the two-dimensional cosmological spacetime. 

In particular, there is no energy conservation in the time-dependent background (\ref{eq:wscft}). 
That is, individual string worldsheet diagrams do not conserve energy and therefore we expect disconnected string diagrams to contribute to a given $n$-point cosmological wavefunction. 
The resulting combinatorics of string diagrams is similar to that of D-instantons in string theory \cite{Polchinski:1994fq,Green:1997tv,Balthazar:2019rnh,Balthazar:2019ypi,Balthazar:2022apu,Sen:2019qqg,Sen:2020cef,Sen:2020eck,Sen:2020ruy,Sen:2021qdk,Sen:2021tpp,Sen:2021jbr,Alexandrov:2021shf,Alexandrov:2021dyl,Alexandrov:2022mmy,Sen:2022clw,Eniceicu:2022nay,Eniceicu:2022dru,Agmon:2022vdj,Eniceicu:2022xvk,Green:2000ke,Billo:2002hm,Billo:2006jm}; however, instead of considering Riemann surfaces with boundaries, in a closed string time-dependent background such as (\ref{eq:wscft}) we consider closed Riemann surfaces. 

A class of disconnected diagrams that factor out of every cosmological wavefunction are empty diagrams, which exponentiate as
\ie
\exp \Big[~ \begin{tikzpicture}[baseline={([yshift=-.5ex]current bounding box.center)},vertex/.style={anchor=base,
    circle,fill=black!25,minimum size=18pt,inner sep=2pt}]
\draw [thick, fill=black!30] (0,0) circle (0.4);
\draw [thick] (-0.4,0) to [out=-35,in=-145] (0.4,0);
\draw [thick, dashed] (-0.4,0) to [out=15,in=165] (0.4,0);
\end{tikzpicture} 
~ + ~
\begin{tikzpicture}[baseline={([yshift=-.5ex]current bounding box.center)},vertex/.style={anchor=base,
    circle,fill=black!25,minimum size=18pt,inner sep=2pt}]
\draw [thick, fill=black!30] (-0.4,0) to [out=85,in=95] (0.4,0) to [out=-95,in=-85] (-0.4,0);
\filldraw [color=white, fill=white] (-0.15,0.01) to [out=40,in=130] (0.15,0.01) to [out=-160,in=-20] (-0.15,0.01);
\draw [thick] (-0.2,0.05) to [out=-40,in=-130] (0.2,0.05);
\draw [thick] (-0.15,0.01) to [out=40,in=130] (0.15,0.01);
\end{tikzpicture}
~ + ~ \cdots  ~ \Big].
\label{eq:emptydiags}
\fe
Each diagram in the exponential is weighted by the Euler character of the two-dimensional worldsheet surface. For instance, the leading contribution coming from the exponentiated empty sphere diagram is expected to give rise to an overall suppression factor of the form $e^{- \frac{A}{g_s^2}}$, where $A$ is a constant (possibly zero). 
The exponentiation of the empty torus diagram, which scales as $g_s^0$, gives rise to a constant $C$. 
More generally an empty Riemann surface of genus $g$ gives a contribution of order $g_s^{2g-2}$. 
While it would be very interesting to calculate the empty sphere\footnote{See \cite{Anninos:2021ene} for related work in this direction.} and torus diagrams in the string theory background (\ref{eq:wscft}), we will not attempt such a computation in this paper. 

The main goal of this paper is to compute the torus one-point string diagram in the cosmological background (\ref{eq:wscft}), 
\ie
\Psi^{g=1}_{\text{1-pt}}(\omega) ~ \equiv ~ 
\begin{tikzpicture}[baseline={([yshift=-.5ex]current bounding box.center)},vertex/.style={anchor=base,
    circle,fill=black!25,minimum size=18pt,inner sep=2pt}]
\draw [thick, fill=black!30] (-0.6,0) to [out=85,in=95] (0.4,0) to [out=-95,in=-85] (-0.6,0);
\filldraw [color=white, fill=white] (-0.15,0.01) to [out=40,in=130] (0.15,0.01) to [out=-160,in=-20] (-0.15,0.01);
\draw [thick] (-0.2,0.05) to [out=-40,in=-130] (0.2,0.05);
\draw [thick] (-0.15,0.01) to [out=40,in=130] (0.15,0.01);
\draw node[cross=3pt, very thick] at (-0.375,0) {};
\end{tikzpicture} ~,
\label{eq:mydiagtorus1pt}
\fe
where the vertex operator insertion is (\ref{eq:vertexop}). 
We will calculate $\Psi^{g=1}_{\text{1-pt}}(\omega)$ numerically by decomposing the one-point functions of the vertex operators on the RHS of (\ref{eq:vertexop}) in $c=1$ Liouville CFT and in $c=25$ Liouville CFT, respectively, into torus one-point Virasoro conformal blocks and integrating over the fundamental domain of the torus moduli space, as described in Section \ref{sec:torus1pt}. 
Although this string theoretic computation is considerably complicated, we will nonetheless find a strikingly simple result for $\Psi^{g=1}_{\text{1-pt}}(\omega)$ as a function of the external closed string energy $\omega$. 

This torus one-point diagram is expected to contribute, for instance, to the full one-point cosmological wavefunction $\Psi(\omega)$ at order $g_s$ (not including the overall scaling implied by the exponentiated sphere diagram) via the following diagrammatic\footnote{In this paper, we will denote by $\Psi^{g}_{n\text{-pt}}(\omega_1,\ldots,\omega_n)$ the $n$-point string \emph{diagram} at genus $g$, and denote by $\Psi(\omega_1,\ldots,\omega_n)$ the full cosmological wavefunction obtained by the sum over disconnected diagrams.}, whose spacetime interpretation is depicted in Figure \ref{fig:spacetime}, 
\ie
\Psi(\omega) ~ \supset ~ & 
\exp \Big[\, \begin{tikzpicture}[baseline={([yshift=-.5ex]current bounding box.center)},vertex/.style={anchor=base,
    circle,fill=black!25,minimum size=18pt,inner sep=2pt}]
\draw [thick, fill=black!30] (0,0) circle (0.4);
\draw [thick] (-0.4,0) to [out=-35,in=-145] (0.4,0);
\draw [thick, dashed] (-0.4,0) to [out=15,in=165] (0.4,0);
\end{tikzpicture} \,\Big] 
\exp \Big[\, \begin{tikzpicture}[baseline={([yshift=-.5ex]current bounding box.center)},vertex/.style={anchor=base,
    circle,fill=black!25,minimum size=18pt,inner sep=2pt}]
\draw [thick, fill=black!30] (-0.4,0) to [out=85,in=95] (0.4,0) to [out=-95,in=-85] (-0.4,0);
\filldraw [color=white, fill=white] (-0.15,0.01) to [out=40,in=130] (0.15,0.01) to [out=-160,in=-20] (-0.15,0.01);
\draw [thick] (-0.2,0.05) to [out=-40,in=-130] (0.2,0.05);
\draw [thick] (-0.15,0.01) to [out=40,in=130] (0.15,0.01);
\end{tikzpicture} \,\Big] ~  
\begin{tikzpicture}[baseline={([yshift=-.5ex]current bounding box.center)},vertex/.style={anchor=base,
    circle,fill=black!25,minimum size=18pt,inner sep=2pt}]
\draw [thick, fill=black!30] (-0.6,0) to [out=85,in=95] (0.4,0) to [out=-95,in=-85] (-0.6,0);
\filldraw [color=white, fill=white] (-0.15,0.01) to [out=40,in=130] (0.15,0.01) to [out=-160,in=-20] (-0.15,0.01);
\draw [thick] (-0.2,0.05) to [out=-40,in=-130] (0.2,0.05);
\draw [thick] (-0.15,0.01) to [out=40,in=130] (0.15,0.01);
\draw node[cross=3pt, very thick] at (-0.375,0) {};
\end{tikzpicture} ~,
\label{eq:full1ptwavefdiag}
\fe
and to the full four-point cosmological wavefunction $\Psi(\omega_1,\omega_2,\omega_3,\omega_4)$ at order $g_s^2$ through the disconnected product of the torus one-point diagram and the sphere three-point diagram,
\ie
\Psi(\omega_1,\omega_2,\omega_3,\omega_4) ~ \supset ~ & 
\exp \Big[\, \begin{tikzpicture}[baseline={([yshift=-.5ex]current bounding box.center)},vertex/.style={anchor=base,
    circle,fill=black!25,minimum size=18pt,inner sep=2pt}]
\draw [thick, fill=black!30] (0,0) circle (0.4);
\draw [thick] (-0.4,0) to [out=-35,in=-145] (0.4,0);
\draw [thick, dashed] (-0.4,0) to [out=15,in=165] (0.4,0);
\end{tikzpicture} \,\Big] 
\exp \Big[\, \begin{tikzpicture}[baseline={([yshift=-.5ex]current bounding box.center)},vertex/.style={anchor=base,
    circle,fill=black!25,minimum size=18pt,inner sep=2pt}]
\draw [thick, fill=black!30] (-0.4,0) to [out=85,in=95] (0.4,0) to [out=-95,in=-85] (-0.4,0);
\filldraw [color=white, fill=white] (-0.15,0.01) to [out=40,in=130] (0.15,0.01) to [out=-160,in=-20] (-0.15,0.01);
\draw [thick] (-0.2,0.05) to [out=-40,in=-130] (0.2,0.05);
\draw [thick] (-0.15,0.01) to [out=40,in=130] (0.15,0.01);
\end{tikzpicture} \,\Big] 
\Big( \begin{tikzpicture}[baseline={([yshift=-.5ex]current bounding box.center)},vertex/.style={anchor=base,
    circle,fill=black!25,minimum size=18pt,inner sep=2pt}]
\draw [thick, fill=black!30] (-0.6,0) to [out=85,in=95] (0.4,0) to [out=-95,in=-85] (-0.6,0);
\filldraw [color=white, fill=white] (-0.15,0.01) to [out=40,in=130] (0.15,0.01) to [out=-160,in=-20] (-0.15,0.01);
\draw [thick] (-0.2,0.05) to [out=-40,in=-130] (0.2,0.05);
\draw [thick] (-0.15,0.01) to [out=40,in=130] (0.15,0.01);
\draw node[cross=3pt, very thick] at (-0.375,0) {};
\node [left] at (-0.5,0) {$_1$};
\end{tikzpicture}
\,\cdot 
\begin{tikzpicture}[baseline={([yshift=-.5ex]current bounding box.center)},vertex/.style={anchor=base,
    circle,fill=black!25,minimum size=18pt,inner sep=2pt}]
\draw [thick, fill=black!30] (0,0) circle (0.4);
\draw [thick] (-0.4,0) to [out=-35,in=-145] (0.4,0);
\draw [thick, dashed] (-0.4,0) to [out=15,in=165] (0.4,0);
\draw node[cross=3pt, very thick] at (-0.2,0.175) {};
\node [above] at (-0.35,0.175) {$_2$};
\draw node[cross=3pt, very thick] at (0.2,0.175) {};
\node [above] at (0.35,0.175) {$_3$};
\draw node[cross=3pt, very thick] at (0,-0.25) {};
\node [below] at (0,-0.3) {$_4$};
\end{tikzpicture}
\,+\, \text{3 permutations \Big)}.
\label{eq:full4ptwavefdiagram}
\fe

The numerical calculation of the torus one-point diagram (\ref{eq:mydiagtorus1pt}) is presented in the next section, followed by a more detailed discussion of the diagrammatics of cosmological wavefunctions and the contribution of (\ref{eq:mydiagtorus1pt}) to the full cosmological wavefunctions (\ref{eq:full1ptwavefdiag}) and (\ref{eq:full4ptwavefdiagram}). 

\section{Calculation of the torus one-point diagram in string perturbation theory}
\label{sec:torus1pt}

In string perturbation theory, the torus one-point diagram in the two-dimensional string cosmology background is given by \cite{Polchinski:1998rq} 
\ie
\Psi^{g=1}_{\text{1-pt}}(\omega) &= 
\frac{(2\pi)^2}{2} \int_{F_0} d^2\tau \, \left\langle b\widetilde{b} c \widetilde{c} \cV_{\omega}(0) \vphantom{\Vhat} \right\rangle_{T^2(\tau)} \\ 
& = g_s C_{T^2} \frac{(2\pi)^2}{2} \int_{F_0} d^2\tau \, |\eta(\tau)|^4 \left\langle \Vhat_{-i\omega}(0) \right\rangle^{T^2(\tau)}_{c=1 \text{ Liouv.}} \left\langle V_{\omega}(0) \vphantom{\Vhat} \right\rangle^{T^2(\tau)}_{c=25 \text{ Liouv.}}
\label{eq:torus1pt1}
\fe
where $F_0 = \{\tau\in\bC |-{1\over 2} \leq {\rm Re}\,\tau \leq {1\over 2}, |\tau|\geq 1 \}$ is the fundamental domain of the torus moduli space\footnote{In our conventions, $d^2\tau = d\tau_1 d\tau_2$ where $\tau = \tau_1 + i \tau_2$.}, and $C_{T^2}$ is a normalization constant associated with the torus topology. 
In time-independent string perturbation theory, $C_{T^2}$ may be fixed by considering the factorization of torus amplitudes into tree-level amplitudes. 
Since at present we do not know the precise implications of unitarity for string theory in time-dependent backgrounds, we will not attempt to compute this normalization constant in this paper.

The one-point correlation functions on the torus $T^2(\tau)$ with modulus $\tau$ in (\ref{eq:torus1pt1}) admit the following Virasoro conformal block decompositions
\ie
\left\langle V_{P_{\rm ext}}(0) \vphantom{\Vhat}\right\rangle^{T^2(\tau)}_{c=25 \text{ Liouv.}} &= \int_0^\infty \frac{dP}{\pi} C(P_{\rm ext},P,P) \cF_{c=25}(h_{\rm ext};1+P^2|q) \cF_{c=25}(h_{\rm ext};1+P^2|\overline{q}),\\
\left\langle \Vhat_{P_{\rm ext}}(0) \vphantom{\Vhat}\right\rangle^{T^2(\tau)}_{c=1 \text{ Liouv.}} &= \int_\cC \frac{d\Phat}{2\pi} \Chat(P_{\rm ext},\Phat,\Phat) \cF_{c=1}(\hhat_{\rm ext};\Phat^2|q) \cF_{c=1}(\hhat_{\rm ext};\Phat^2|\overline{q}),
\label{eq:torus1pts}
\fe
where $\cF_{c}(h_{\rm ext};h_{\rm int}|q)$ is the holomorphic torus one-point Virasoro conformal block at central charge $c$ with external weight $h_{\rm ext}$ and internal weight $h_{\rm int}$, evaluated at a value of the parameter $q=e^{2\pi i \tau}$ where $\tau$ is the modulus of the torus. 
In (\ref{eq:torus1pts}), $h_{\rm ext} = 1+P_{\rm ext}^2$ and $\hhat_{\rm ext} = P_{\rm ext}^2$. 
The three-point function coefficients $C(P_1,P_2,P_3)$ and $\Chat(P_1,P_2,P_3)$ appearing in (\ref{eq:torus1pts}) were reviewed in \cite{Rodriguez:2023kkl}, which we recall here for convenience  \cite{Dorn:1994xn,Zamolodchikov:1995aa,Ribault:2015sxa,Schomerus:2003vv,Kostov:2005kk,Zamolodchikov:2005fy,Harlow:2011ny,McElgin:2007ak,Giribet:2011zx}
\ie
C(P_1,P_2,P_3) &= \frac{1}{\ups(1+i(P_1+P_2+P_3))} \left[ \frac{2P_1\ups(1+2iP_1)}{\ups(1+i(P_2+P_3-P_1))}\times(\text{2 permutations}) \right], \\
\Chat(P_1,P_2,P_3) &= \ups(1+P_1+P_2+P_3) \left[ \frac{\ups(1+P_2+P_3-P_1)}{\ups(1+2P_1)}\times(\text{2 permutations}) \right], \vphantom{\int^{\Vhat}}
\label{eq:DOZZs}
\fe
where the function $\ups(x)$ is defined as
\ie
\ups(x) = \frac{1}{\Gamma_1(x)\Gamma_1(2-x)},
\label{eq:defUps1}
\fe
where $\Gamma_1(x)$ is related to the Barnes G-function $G(x)$ by $\Gamma_1(x) = (2\pi)^{(x-1)/2}(G(x))^{-1}$ \cite{Balthazar:2018qdv}.

The contour of integration $\cC$ over the intermediate states with Liouville momentum $\Phat$ in the $c=1$
 torus one-point function (\ref{eq:torus1pts}) is chosen to be parallel to the real axis but shifted a small amount in the imaginary axis in order to avoid the poles of the $c=1$ Liouville three-point function coefficient (\ref{eq:DOZZs}) and of the $c=1$ conformal blocks that reside on the real axis at half-integer values \cite{Ribault:2015sxa,Bautista:2019jau,Rodriguez:2023kkl}.
Using the modular covariance relations (\ref{eq:modcov}) one can verify that the integrand times the measure in (\ref{eq:torus1pt1}) is modular invariant. 

The tours one-point Virasoro conformal block $\cF_{c}(h_{\rm ext};h_{\rm int}|q)$ can be expressed as \cite{Hadasz:2009db,Cho:2017oxl}
\ie
\cF_{c}(h_{\rm ext};h_{\rm int}|q) = q^{h_{\rm int} - c/24} \left( \prod_{m=1}^{\infty} \frac{1}{1-q^m} \right) \mathcal{H}_c(h_{\rm ext};h_{\rm int}|q),
\label{eq:torusCBelliptic}
\fe
where the so-called elliptic conformal block $\mathcal{H}_c(h_{\rm ext};h_{\rm int}|q)$ admits a power series expansion in $q=e^{2\pi i\tau}$ that starts at $1$ and that can be computed efficiently with a recursion relation in the internal weight $h_{\rm int}$, as briefly reviewed in Appendix \ref{sec:torusblocks}.
Decomposing the Liouville one-point functions in (\ref{eq:torus1pt1}) into Virasoro conformal blocks and making use of (\ref{eq:torusCBelliptic}) we obtain that the torus one-point diagram in the two-dimensional string cosmology background takes the form,
\ie
\Psi^{g=1}_{\text{1-pt}}(\omega) = g_s C_{T^2} \frac{(2\pi)^2}{2}& \int_{F_0} d^2\tau \int_\cC \frac{d\Phat}{2\pi} \Chat(-i\omega,\Phat,\Phat) |q|^{2\Phat^2} \cH_{c=1}(\hhat; \Phat^2|q) \cH_{c=1}(\hhat; \Phat^2|\overline{q}) \\
\times & \int_0^\infty\frac{dP}{\pi} C(\omega,P,P) |q|^{2P^2} \cH_{c=25}(h; 1+P^2|q) \cH_{c=25}(h; 1+P^2|\overline{q}),
\label{eq:torus1ptfinal}
\fe
where $h=1+\omega^2$ and $\hhat=-\omega^2$.

Let us analyze the behavior of (\ref{eq:torus1ptfinal}) near the cusp $\tau_2 \to \infty$ of the fundamental domain, where we parametrize the torus modulus by $\tau = \tau_1 + i \tau_2$.
Since to leading order at large $\tau_2$ the torus one-point elliptic conformal blocks $\cH_c(h_{\rm ext};h_{\rm int}|q) \simeq 1$, the moduli integral of (\ref{eq:torus1ptfinal}) behaves as
\ie
\int^{\infty}d\tau_2 \biggl( \int_\cC \frac{d\Phat}{2\pi} \Chat(-i\omega,\Phat,\Phat) e^{-4\pi\tau_2\Phat^2} \biggr) \biggl( \int_0^\infty\frac{dP}{\pi} C(\omega,P,P) e^{-4\pi\tau_2 P^2} \biggr)
\label{eq:cusp1}
\fe
In the limit that $\tau_2\to\infty$, the integrals over the intermediate Liouville momenta are dominated by their values near $P=0$ and $\Phat=0$. Using Laplace's method, we can write an asymptotic expansion at large $\tau_2$ as
\ie
\int_\cC \frac{d\Phat}{2\pi} \Chat(-i\omega,\Phat,\Phat) e^{-4\pi\tau_2\Phat^2} &\simeq \sum_{n=0}^{\infty}\Chat^{(n)}(-i\omega,0,0) \frac{2^{-2(n+1)}\pi^{-\frac{n+2}{2}}}{\Gamma({n\over 2} + 1)} \, \tau_2^{-\frac{n+1}{2}},\\
\int_0^\infty\frac{dP}{\pi} C(\omega,P,P) e^{-4\pi\tau_2 P^2} &\simeq \sum_{n=0}^{\infty}C^{(n)}(\omega,0,0) \frac{2^{-2(n+1)}\pi^{-\frac{n+2}{2}}}{\Gamma({n\over 2} + 1)} \, \tau_2^{-\frac{n+1}{2}},
\label{eq:asymexp}
\fe
where $\Chat^{(n)}(-i\omega,0,0)$ and $C^{(n)}(\omega,0,0)$ denote the $n$-th derivative of $\Chat(-i\omega,\Phat,\Phat)$ and $C(\omega,P,P)$ with respect to $\Phat$ and $P$, respectively, and evaluated at $\Phat=0$ and $P=0$. 
Taking the first nonzero terms in (\ref{eq:asymexp}), we obtain that the moduli integral (\ref{eq:cusp1}) behaves as 
\ie
\frac{\omega}{16\pi^3}\int^{\infty} d\tau_2 \,\, \tau_2^{-2},
\label{eq:nodiv}
\fe
and is therefore convergent. 
Hence, the string one-point diagram (\ref{eq:torus1ptfinal}) requires no regularization (or definition via analytic continuation) in contrast to the usual case of string S-matrix elements in time-independent string perturbation theory.

In the numerical evaluation of (\ref{eq:torus1ptfinal}), we will employ the following strategy. We first split the fundamental domain $F_0$ of the torus modulus into two regions: (I) $\tau\in F_0$ with $\tau_2 \leq \tau_2^{\rm max}$, and (II) $\tau\in F_0$ with $\tau_2 \geq \tau_2^{\rm max}$, for a sufficiently large value of $\tau_2^{\rm max}$. 
In region (I), we first perform the integrals over the intermediate Liouville momenta $\Phat$ and $P$ separately and for a fixed value of $\tau$. These two integrations are performed numerically with truncated conformal blocks $\cH_{c=1}(\hhat; \Phat^2|q)$ and $\cH_{c=25}(h; 1+P^2|q)$ up to order $q^8$. We then perform the integration over $\tau$ in region (I) numerically.
In region (II), we may approximate the moduli integrand by the expressions in (\ref{eq:cusp1}) and (\ref{eq:asymexp}). We include a sufficient number of terms in the asymptotic expansions (\ref{eq:asymexp}) such that the resulting moduli integral is accurate to order $(\tau_2^{\rm max})^{-6}$. 

As we approach the limit of the central charges $c\to 1$ and $c\to 25$, the numerical calculation of the torus one-point Virasoro conformal blocks $\cH_c(h_{\rm ext};h_{\rm int}|q)$ through the recursion relation (\ref{eq:recrel}) involves delicate cancelations. In order to avoid loss of precision, we evaluate the conformal blocks for values of the central charges sufficiently close but not exactly equal to 1 and 25. 
In summary, in the numerical calculation of (\ref{eq:torus1ptfinal}) we set the values for the central charges $c = 1 - 10^{-5}$ and $c=25+10^{-6}$, parametrized the contour of integration $\cC$ over the intermediate $c=1$ Liouville momentum by $\Phat = p + i\epsilon$ with $p\in\bR$ and $\epsilon=10^{-1}$, and set $\tau_2^{\rm max} = 15$ in the splitting of the fundamental domain $F_0$ described in the previous paragraph.

\begin{figure}[h!]
\centering

\includegraphics[width=0.8\textwidth]{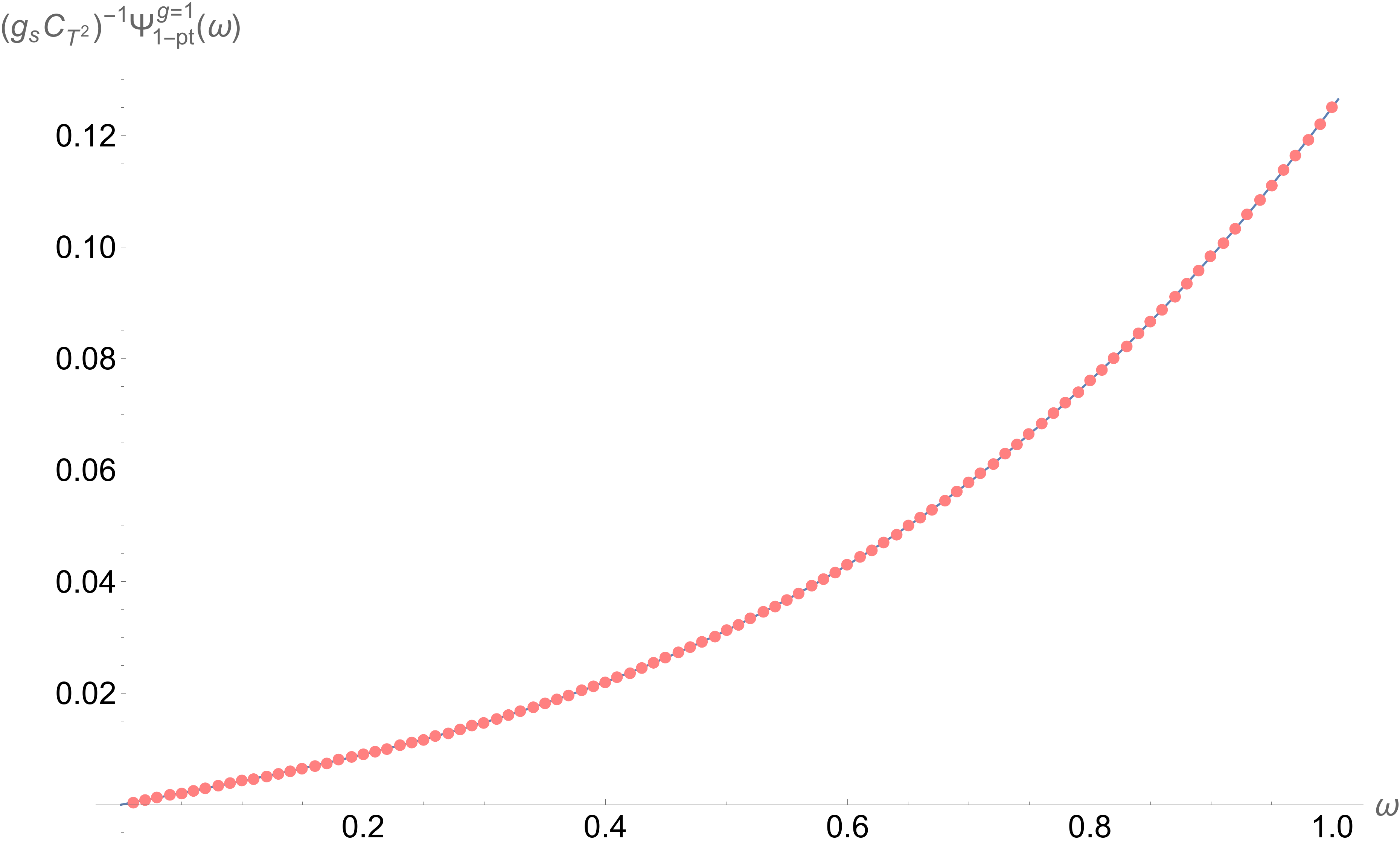}

\caption{Shown in dots are the numerical results for the torus one-point diagram (\ref{eq:torus1ptfinal}) in two-dimensional string cosmology, for a range of energies $\omega\in[0,1]$ of the outgoing asymptotic closed string state. The fit (\ref{eq:myfit}) is shown in the solid curve.}
\label{fig:data}
\end{figure}

Numerical results for the one-point torus diagram (\ref{eq:torus1ptfinal}), calculated with the strategy outline above and for the choice of outgoing external energy of the closed string $\omega\in [0,1]$, are shown in Figure \ref{fig:data}. 
We find to a very high accuracy that the numerical results fit the following function
\ie
\Psi^{g=1}_{\text{1-pt}}(\omega) = \frac{g_s C_{T^2}}{24} \,\omega \left( 1 + 2\omega^2 \right). \vphantom{\biggl)}
\label{eq:myfit}
\fe
The largest discrepancy between the numerical results for (\ref{eq:torus1ptfinal}) and the fit (\ref{eq:myfit}) is $10^{-5}\,\%$.

\section{Diagrammatics of cosmological wavefunctions and discussion of results}
\label{sec:discussion}

In this paper, we have explicitly computed the torus one-point diagram in the two-dimensional string cosmology background considered in \cite{Rodriguez:2023kkl}.
The main result of this paper is the conjectural result (\ref{eq:myfit}), verified numerically to very high accuracy. 

As mentioned in the introduction, the torus one-point diagram is expected to contribute to the full cosmological one-point wavefunction $\Psi(\omega)$, which takes the form
\ie
\Psi(\omega) \, & = \,   
\exp \Big[\, \begin{tikzpicture}[baseline={([yshift=-.5ex]current bounding box.center)},vertex/.style={anchor=base,
    circle,fill=black!25,minimum size=18pt,inner sep=2pt}]
\draw [thick, fill=black!30] (0,0) circle (0.4);
\draw [thick] (-0.4,0) to [out=-35,in=-145] (0.4,0);
\draw [thick, dashed] (-0.4,0) to [out=15,in=165] (0.4,0);
\end{tikzpicture} \,\Big] 
\exp \Big[\, \begin{tikzpicture}[baseline={([yshift=-.5ex]current bounding box.center)},vertex/.style={anchor=base,
    circle,fill=black!25,minimum size=18pt,inner sep=2pt}]
\draw [thick, fill=black!30] (-0.4,0) to [out=85,in=95] (0.4,0) to [out=-95,in=-85] (-0.4,0);
\filldraw [color=white, fill=white] (-0.15,0.01) to [out=40,in=130] (0.15,0.01) to [out=-160,in=-20] (-0.15,0.01);
\draw [thick] (-0.2,0.05) to [out=-40,in=-130] (0.2,0.05);
\draw [thick] (-0.15,0.01) to [out=40,in=130] (0.15,0.01);
\end{tikzpicture} \,\Big] \Big\{ ~ 
\begin{tikzpicture}[baseline={([yshift=-.5ex]current bounding box.center)},vertex/.style={anchor=base,
    circle,fill=black!25,minimum size=18pt,inner sep=2pt}]
\draw [thick, fill=black!30] (0,0) circle (0.4);
\draw [thick] (-0.4,0) to [out=-35,in=-145] (0.4,0);
\draw [thick, dashed] (-0.4,0) to [out=15,in=165] (0.4,0);
\draw node[cross=3pt, very thick] at (0,0.2) {};
\end{tikzpicture}
~ + \, \Big(\, 
\begin{tikzpicture}[baseline={([yshift=-.5ex]current bounding box.center)},vertex/.style={anchor=base,
    circle,fill=black!25,minimum size=18pt,inner sep=2pt}]
\draw [thick, fill=black!30] (-0.6,0) to [out=85,in=95] (0.4,0) to [out=-95,in=-85] (-0.6,0);
\filldraw [color=white, fill=white] (-0.15,0.01) to [out=40,in=130] (0.15,0.01) to [out=-160,in=-20] (-0.15,0.01);
\draw [thick] (-0.2,0.05) to [out=-40,in=-130] (0.2,0.05);
\draw [thick] (-0.15,0.01) to [out=40,in=130] (0.15,0.01);
\draw node[cross=3pt, very thick] at (-0.375,0) {};
\end{tikzpicture} 
~ + ~ 
\begin{tikzpicture}[baseline={([yshift=-.5ex]current bounding box.center)},vertex/.style={anchor=base,
    circle,fill=black!25,minimum size=18pt,inner sep=2pt}]
\draw [thick, fill=black!30] (0,0) circle (0.4);
\draw [thick] (-0.4,0) to [out=-35,in=-145] (0.4,0);
\draw [thick, dashed] (-0.4,0) to [out=15,in=165] (0.4,0);
\draw node[cross=3pt, very thick] at (0,0.2) {};
\end{tikzpicture} 
\cdot
\begin{tikzpicture}[baseline={([yshift=-.5ex]current bounding box.center)},vertex/.style={anchor=base,
    circle,fill=black!25,minimum size=18pt,inner sep=2pt}]
\draw [thick, fill=black!30] (-0.4,0) to [out=85,in=170] (0.4,0.15) to [out=10,in=95] (1.2,0) to [out=-95,in=-10] (0.4,-0.15) to [out=-170,in=-85] (-0.4,0);
\filldraw [color=white, fill=white] (-0.15,0.01) to [out=40,in=130] (0.15,0.01) to [out=-160,in=-20] (-0.15,0.01);
\draw [thick] (-0.2,0.05) to [out=-40,in=-130] (0.2,0.05);
\draw [thick] (-0.15,0.01) to [out=40,in=130] (0.15,0.01);
\filldraw [color=white, fill=white] (0.65,0.01) to [out=40,in=130] (0.95,0.01) to [out=-160,in=-20] (0.65,0.01);
\draw [thick] (0.6,0.05) to [out=-40,in=-130] (1,0.05);
\draw [thick] (0.65,0.01) to [out=40,in=130] (0.95,0.01);
\end{tikzpicture}
~ \Big) \, + ~ \mathcal{O}(g_s^3) \Big\} \\
& = \, e^{-{A\over g_s^2}} \, C  \,
\Big\{ ~ 
\begin{tikzpicture}[baseline={([yshift=-.5ex]current bounding box.center)},vertex/.style={anchor=base,
    circle,fill=black!25,minimum size=18pt,inner sep=2pt}]
\draw [thick, fill=black!30] (0,0) circle (0.4);
\draw [thick] (-0.4,0) to [out=-35,in=-145] (0.4,0);
\draw [thick, dashed] (-0.4,0) to [out=15,in=165] (0.4,0);
\draw node[cross=3pt, very thick] at (0,0.2) {};
\end{tikzpicture}
~ + \, \Big(\, 
\begin{tikzpicture}[baseline={([yshift=-.5ex]current bounding box.center)},vertex/.style={anchor=base,
    circle,fill=black!25,minimum size=18pt,inner sep=2pt}]
\draw [thick, fill=black!30] (-0.6,0) to [out=85,in=95] (0.4,0) to [out=-95,in=-85] (-0.6,0);
\filldraw [color=white, fill=white] (-0.15,0.01) to [out=40,in=130] (0.15,0.01) to [out=-160,in=-20] (-0.15,0.01);
\draw [thick] (-0.2,0.05) to [out=-40,in=-130] (0.2,0.05);
\draw [thick] (-0.15,0.01) to [out=40,in=130] (0.15,0.01);
\draw node[cross=3pt, very thick] at (-0.375,0) {};
\end{tikzpicture} 
~ + ~ 
\begin{tikzpicture}[baseline={([yshift=-.5ex]current bounding box.center)},vertex/.style={anchor=base,
    circle,fill=black!25,minimum size=18pt,inner sep=2pt}]
\draw [thick, fill=black!30] (0,0) circle (0.4);
\draw [thick] (-0.4,0) to [out=-35,in=-145] (0.4,0);
\draw [thick, dashed] (-0.4,0) to [out=15,in=165] (0.4,0);
\draw node[cross=3pt, very thick] at (0,0.2) {};
\end{tikzpicture} 
\cdot
\begin{tikzpicture}[baseline={([yshift=-.5ex]current bounding box.center)},vertex/.style={anchor=base,
    circle,fill=black!25,minimum size=18pt,inner sep=2pt}]
\draw [thick, fill=black!30] (-0.4,0) to [out=85,in=170] (0.4,0.15) to [out=10,in=95] (1.2,0) to [out=-95,in=-10] (0.4,-0.15) to [out=-170,in=-85] (-0.4,0);
\filldraw [color=white, fill=white] (-0.15,0.01) to [out=40,in=130] (0.15,0.01) to [out=-160,in=-20] (-0.15,0.01);
\draw [thick] (-0.2,0.05) to [out=-40,in=-130] (0.2,0.05);
\draw [thick] (-0.15,0.01) to [out=40,in=130] (0.15,0.01);
\filldraw [color=white, fill=white] (0.65,0.01) to [out=40,in=130] (0.95,0.01) to [out=-160,in=-20] (0.65,0.01);
\draw [thick] (0.6,0.05) to [out=-40,in=-130] (1,0.05);
\draw [thick] (0.65,0.01) to [out=40,in=130] (0.95,0.01);
\end{tikzpicture}
~ \Big) \, + ~ \mathcal{O}(g_s^3) \Big\},
\label{eq:full1ptwavef}
\fe
where $A$ and $C$ are constants, which we have not attempted to compute in this paper. 
For instance, our result (\ref{eq:myfit}) implies that the contribution of the torus one-point diagram $\Psi^{g=1}_{1\text{-pt}}(\omega)$ to the full cosmological wavefunction (\ref{eq:full1ptwavef}) at order $g_s$ is given by
\ie
\Psi(\omega) ~ \supset ~  
\exp \Big[\, \begin{tikzpicture}[baseline={([yshift=-.5ex]current bounding box.center)},vertex/.style={anchor=base,
    circle,fill=black!25,minimum size=18pt,inner sep=2pt}]
\draw [thick, fill=black!30] (0,0) circle (0.4);
\draw [thick] (-0.4,0) to [out=-35,in=-145] (0.4,0);
\draw [thick, dashed] (-0.4,0) to [out=15,in=165] (0.4,0);
\end{tikzpicture} \,\Big] 
\exp \Big[\, \begin{tikzpicture}[baseline={([yshift=-.5ex]current bounding box.center)},vertex/.style={anchor=base,
    circle,fill=black!25,minimum size=18pt,inner sep=2pt}]
\draw [thick, fill=black!30] (-0.4,0) to [out=85,in=95] (0.4,0) to [out=-95,in=-85] (-0.4,0);
\filldraw [color=white, fill=white] (-0.15,0.01) to [out=40,in=130] (0.15,0.01) to [out=-160,in=-20] (-0.15,0.01);
\draw [thick] (-0.2,0.05) to [out=-40,in=-130] (0.2,0.05);
\draw [thick] (-0.15,0.01) to [out=40,in=130] (0.15,0.01);
\end{tikzpicture} \,\Big] ~  
\begin{tikzpicture}[baseline={([yshift=-.5ex]current bounding box.center)},vertex/.style={anchor=base,
    circle,fill=black!25,minimum size=18pt,inner sep=2pt}]
\draw [thick, fill=black!30] (-0.6,0) to [out=85,in=95] (0.4,0) to [out=-95,in=-85] (-0.6,0);
\filldraw [color=white, fill=white] (-0.15,0.01) to [out=40,in=130] (0.15,0.01) to [out=-160,in=-20] (-0.15,0.01);
\draw [thick] (-0.2,0.05) to [out=-40,in=-130] (0.2,0.05);
\draw [thick] (-0.15,0.01) to [out=40,in=130] (0.15,0.01);
\draw node[cross=3pt, very thick] at (-0.375,0) {};
\end{tikzpicture}
~\, = ~\, e^{-{A\over g_s^2}} \, C  \, \frac{g_s C_{T^2}}{24} \,\omega \left( 1 + 2\omega^2 \right).
\label{eq:1ptwavefcont1}
\fe

%
%

While it is tempting to assume that the one-point string diagram on the sphere vanishes and hence (\ref{eq:1ptwavefcont1}) is the entire contribution to the full cosmological one-point wavefunction up to order\footnote{Not including the overall factor of $e^{-{A\over g_s^2}}$ from the exponentiated empty sphere diagram, for simplicity.} $g_s$, there are string theory backgrounds in which the sphere one-point diagram has been argued to be nonzero for certain classes of vertex operators (see for example \cite{Seiberg:1990eb,Dorn:1994xn,Kutasov:1999xu,Troost:2011ud}). 
Nevertheless, a naive argument for the vanishing of the sphere one-point string diagram for an on-shell state (\ref{eq:vertexop}) with a nonzero energy $\omega\neq 0$ is that the one-point functions of the individual fixed vertex operators in $c=1$ and $c=25$ Liouville CFT, respectively, vanish by scaling symmetry and cannot compensate the division by the infinite volume of the residual conformal Killing group of the sphere, after fixing the position of the full on-shell vertex operator (\ref{eq:vertexop}).

A more physical argument to expect the vanishing of the string one-point diagram on the sphere is as follows. The cosmological wavefunction $\Psi(\omega_1,\ldots,\omega_n)$ has the physical interpretation as a measure of the production of an $n$-particle state in the infinite future. A nonzero one-point string diagram would imply a production rate that is \emph{enhanced} by a factor of $g_s^{-n}$ from the leading order contribution to the cosmological wavefunction $\Psi(\omega_1,\ldots,\omega_n)$ of a disconnected product of $n$ sphere one-point diagrams.
On the other hand, if we assume that the one-point string diagram vanishes, it is easy to convince oneself that the leading order contribution to a cosmological wavefunction $\Psi(\omega_1,\ldots,\omega_n)$ is of order $g_s^0$, if it involves the sphere two-point diagram, or of order $g_s$, yielding a healthier-looking genus expansion. 

For instance, if we assume the sphere one-point diagram vanishes, the leading order contribution to the full two-, three-, and four-point cosmological wavefunctions is given by
\ie
\Psi(\omega_1,\omega_2) & = 
\exp \Big[\, \begin{tikzpicture}[baseline={([yshift=-.5ex]current bounding box.center)},vertex/.style={anchor=base,
    circle,fill=black!25,minimum size=18pt,inner sep=2pt}]
\draw [thick, fill=black!30] (0,0) circle (0.4);
\draw [thick] (-0.4,0) to [out=-35,in=-145] (0.4,0);
\draw [thick, dashed] (-0.4,0) to [out=15,in=165] (0.4,0);
\end{tikzpicture} \,\Big] 
\exp \Big[\, \begin{tikzpicture}[baseline={([yshift=-.5ex]current bounding box.center)},vertex/.style={anchor=base,
    circle,fill=black!25,minimum size=18pt,inner sep=2pt}]
\draw [thick, fill=black!30] (-0.4,0) to [out=85,in=95] (0.4,0) to [out=-95,in=-85] (-0.4,0);
\filldraw [color=white, fill=white] (-0.15,0.01) to [out=40,in=130] (0.15,0.01) to [out=-160,in=-20] (-0.15,0.01);
\draw [thick] (-0.2,0.05) to [out=-40,in=-130] (0.2,0.05);
\draw [thick] (-0.15,0.01) to [out=40,in=130] (0.15,0.01);
\end{tikzpicture} \,\Big] 
\Big\{ 
\begin{tikzpicture}[baseline={([yshift=-.5ex]current bounding box.center)},vertex/.style={anchor=base,
    circle,fill=black!25,minimum size=18pt,inner sep=2pt}]
\draw [thick, fill=black!30] (0,0) circle (0.4);
\draw [thick] (-0.4,0) to [out=-35,in=-145] (0.4,0);
\draw [thick, dashed] (-0.4,0) to [out=15,in=165] (0.4,0);
\draw node[cross=3pt, very thick] at (-0.2,0.175) {};
\node [above] at (-0.35,0.175) {$_1$};
\draw node[cross=3pt, very thick] at (0.2,0.175) {};
\node [above] at (0.35,0.175) {$_2$};
\node [below] at (0,-0.3) {$\vphantom{_0}$};
\end{tikzpicture}
~ + ~ \mathcal{O}(g_s^2) \, \Big\}, \\
\Psi(\omega_1,\omega_2,\omega_3) & =  
\exp \Big[\, \begin{tikzpicture}[baseline={([yshift=-.5ex]current bounding box.center)},vertex/.style={anchor=base,
    circle,fill=black!25,minimum size=18pt,inner sep=2pt}]
\draw [thick, fill=black!30] (0,0) circle (0.4);
\draw [thick] (-0.4,0) to [out=-35,in=-145] (0.4,0);
\draw [thick, dashed] (-0.4,0) to [out=15,in=165] (0.4,0);
\end{tikzpicture} \,\Big] 
\exp \Big[\, \begin{tikzpicture}[baseline={([yshift=-.5ex]current bounding box.center)},vertex/.style={anchor=base,
    circle,fill=black!25,minimum size=18pt,inner sep=2pt}]
\draw [thick, fill=black!30] (-0.4,0) to [out=85,in=95] (0.4,0) to [out=-95,in=-85] (-0.4,0);
\filldraw [color=white, fill=white] (-0.15,0.01) to [out=40,in=130] (0.15,0.01) to [out=-160,in=-20] (-0.15,0.01);
\draw [thick] (-0.2,0.05) to [out=-40,in=-130] (0.2,0.05);
\draw [thick] (-0.15,0.01) to [out=40,in=130] (0.15,0.01);
\end{tikzpicture} \,\Big] 
\Big\{
\begin{tikzpicture}[baseline={([yshift=-.5ex]current bounding box.center)},vertex/.style={anchor=base,
    circle,fill=black!25,minimum size=18pt,inner sep=2pt}]
\draw [thick, fill=black!30] (0,0) circle (0.4);
\draw [thick] (-0.4,0) to [out=-35,in=-145] (0.4,0);
\draw [thick, dashed] (-0.4,0) to [out=15,in=165] (0.4,0);
\draw node[cross=3pt, very thick] at (-0.2,0.175) {};
\node [above] at (-0.35,0.175) {$_1$};
\draw node[cross=3pt, very thick] at (0.2,0.175) {};
\node [above] at (0.35,0.175) {$_2$};
\draw node[cross=3pt, very thick] at (0,-0.25) {};
\node [below] at (0,-0.3) {$_3$};
\end{tikzpicture}
+
\Big(
\begin{tikzpicture}[baseline={([yshift=-.5ex]current bounding box.center)},vertex/.style={anchor=base,
    circle,fill=black!25,minimum size=18pt,inner sep=2pt}]
\draw [thick, fill=black!30] (0,0) circle (0.4);
\draw [thick] (-0.4,0) to [out=-35,in=-145] (0.4,0);
\draw [thick, dashed] (-0.4,0) to [out=15,in=165] (0.4,0);
\draw node[cross=3pt, very thick] at (-0.2,0.175) {};
\node [above] at (-0.35,0.175) {$_1$};
\draw node[cross=3pt, very thick] at (0.2,0.175) {};
\node [above] at (0.35,0.175) {$_2$};
\node [below] at (0,-0.3) {$\vphantom{_0}$};
\end{tikzpicture}
\cdot
\begin{tikzpicture}[baseline={([yshift=-.5ex]current bounding box.center)},vertex/.style={anchor=base,
    circle,fill=black!25,minimum size=18pt,inner sep=2pt}]
\draw [thick, fill=black!30] (-0.6,0) to [out=85,in=95] (0.4,0) to [out=-95,in=-85] (-0.6,0);
\filldraw [color=white, fill=white] (-0.15,0.01) to [out=40,in=130] (0.15,0.01) to [out=-160,in=-20] (-0.15,0.01);
\draw [thick] (-0.2,0.05) to [out=-40,in=-130] (0.2,0.05);
\draw [thick] (-0.15,0.01) to [out=40,in=130] (0.15,0.01);
\draw node[cross=3pt, very thick] at (-0.375,0) {};
\node[above] at (-0.375,0.15) {$_3$};
\node[below] at (-0.375,-0.15) {$\vphantom{_3}$};
\end{tikzpicture}
\,+\, \text{2 perms\Big)}
\, + \, \mathcal{O}(g_s^3) \, \Big\}, \\
\Psi(\omega_1,\omega_2,\omega_3,\omega_4) &=  
\exp \Big[\, \begin{tikzpicture}[baseline={([yshift=-.5ex]current bounding box.center)},vertex/.style={anchor=base,
    circle,fill=black!25,minimum size=18pt,inner sep=2pt}]
\draw [thick, fill=black!30] (0,0) circle (0.4);
\draw [thick] (-0.4,0) to [out=-35,in=-145] (0.4,0);
\draw [thick, dashed] (-0.4,0) to [out=15,in=165] (0.4,0);
\end{tikzpicture} \,\Big] 
\exp \Big[\, \begin{tikzpicture}[baseline={([yshift=-.5ex]current bounding box.center)},vertex/.style={anchor=base,
    circle,fill=black!25,minimum size=18pt,inner sep=2pt}]
\draw [thick, fill=black!30] (-0.4,0) to [out=85,in=95] (0.4,0) to [out=-95,in=-85] (-0.4,0);
\filldraw [color=white, fill=white] (-0.15,0.01) to [out=40,in=130] (0.15,0.01) to [out=-160,in=-20] (-0.15,0.01);
\draw [thick] (-0.2,0.05) to [out=-40,in=-130] (0.2,0.05);
\draw [thick] (-0.15,0.01) to [out=40,in=130] (0.15,0.01);
\end{tikzpicture} \,\Big] 
\Big\{ \frac{1}{2}
\Big(
\begin{tikzpicture}[baseline={([yshift=-.5ex]current bounding box.center)},vertex/.style={anchor=base,
    circle,fill=black!25,minimum size=18pt,inner sep=2pt}]
\draw [thick, fill=black!30] (0,0) circle (0.4);
\draw [thick] (-0.4,0) to [out=-35,in=-145] (0.4,0);
\draw [thick, dashed] (-0.4,0) to [out=15,in=165] (0.4,0);
\draw node[cross=3pt, very thick] at (-0.2,0.175) {};
\node [above] at (-0.35,0.175) {$_1$};
\draw node[cross=3pt, very thick] at (0.2,0.175) {};
\node [above] at (0.35,0.175) {$_2$};
\node [below] at (0,-0.3) {$\vphantom{_0}$};
\end{tikzpicture}
\cdot
\begin{tikzpicture}[baseline={([yshift=-.5ex]current bounding box.center)},vertex/.style={anchor=base,
    circle,fill=black!25,minimum size=18pt,inner sep=2pt}]
\draw [thick, fill=black!30] (0,0) circle (0.4);
\draw [thick] (-0.4,0) to [out=-35,in=-145] (0.4,0);
\draw [thick, dashed] (-0.4,0) to [out=15,in=165] (0.4,0);
\draw node[cross=3pt, very thick] at (-0.2,0.175) {};
\node [above] at (-0.35,0.175) {$_3$};
\draw node[cross=3pt, very thick] at (0.2,0.175) {};
\node [above] at (0.35,0.175) {$_4$};
\node [below] at (0,-0.3) {$\vphantom{_0}$};
\end{tikzpicture}
\,+\, \text{2 perms\Big)}
\, + \, \mathcal{O}(g_s^2) \, \Big\},
\label{eq:fullLOs}
\fe
whose diagrams inside the curly braces start at order $g_s^0$, $g_s^1$, and $g_s^0$, respectively, as well as (\ref{eq:1ptwavefcont1}) for the case of the one-point wavefunction. 

In time-independent string perturbation theory, the two-point scattering amplitude may be deduced from perturbative unitarity of the S-matrix, and in flat space it reduces to the free-particle answer $\langle P | P' \rangle = 2P^0 (2\pi)^{D-1}\delta^{D-1}(\vec{P}-\vec{P}')$ where $P^{\mu}=(P^0,\vec{P})$ obeys $P^2=-M^2$ \cite{Polchinski:1998rq,Erbin:2019uiz}. 
While we have not attempted to compute the two-point sphere diagram in the background (\ref{eq:wscft}), it is natural to expect that it reduces to the free particle case as well, $\Psi^{g=0}_{2{\rm -pt}}(\omega_1,\omega_2) \propto \omega_1\delta(\omega_1-\omega_2)$, as a result of Liouville momentum conservation from the two-point correlation function of vertex operators in the spatial $c=25$ Liouville worldsheet CFT sector. 
In this sense, contributions to a given $n$-point cosmological wavefunction that involve a sphere two-point diagram (e.g. the leading order contributions to the two- and four-point wavefunctions, or the diagrams inside the parenthesis contributing to the three-point wavefunction in (\ref{eq:fullLOs})) may be distinguished by the presence of a delta function setting two of the outgoing energies to be equal. 

For example, the leading order contribution to the three-point cosmological wavefunction that would not involve a delta function coming from a sphere two-point diagram is simply given by the sphere three-point diagram $\Psi^{g=0}_{3\text{-pt}}(\omega_1,\omega_2,\omega_3)$ computed in \cite{Rodriguez:2023kkl},
\ie
\Psi(\omega_1,\omega_2,\omega_3) \supset  
\exp \Big[\, \begin{tikzpicture}[baseline={([yshift=-.5ex]current bounding box.center)},vertex/.style={anchor=base,
    circle,fill=black!25,minimum size=18pt,inner sep=2pt}]
\draw [thick, fill=black!30] (0,0) circle (0.4);
\draw [thick] (-0.4,0) to [out=-35,in=-145] (0.4,0);
\draw [thick, dashed] (-0.4,0) to [out=15,in=165] (0.4,0);
\end{tikzpicture} \,\Big] 
\exp \Big[\, \begin{tikzpicture}[baseline={([yshift=-.5ex]current bounding box.center)},vertex/.style={anchor=base,
    circle,fill=black!25,minimum size=18pt,inner sep=2pt}]
\draw [thick, fill=black!30] (-0.4,0) to [out=85,in=95] (0.4,0) to [out=-95,in=-85] (-0.4,0);
\filldraw [color=white, fill=white] (-0.15,0.01) to [out=40,in=130] (0.15,0.01) to [out=-160,in=-20] (-0.15,0.01);
\draw [thick] (-0.2,0.05) to [out=-40,in=-130] (0.2,0.05);
\draw [thick] (-0.15,0.01) to [out=40,in=130] (0.15,0.01);
\end{tikzpicture} \,\Big] 
\begin{tikzpicture}[baseline={([yshift=-.5ex]current bounding box.center)},vertex/.style={anchor=base,
    circle,fill=black!25,minimum size=18pt,inner sep=2pt}]
\draw [thick, fill=black!30] (0,0) circle (0.4);
\draw [thick] (-0.4,0) to [out=-35,in=-145] (0.4,0);
\draw [thick, dashed] (-0.4,0) to [out=15,in=165] (0.4,0);
\draw node[cross=3pt, very thick] at (-0.2,0.175) {};
\node [above] at (-0.35,0.175) {$_1$};
\draw node[cross=3pt, very thick] at (0.2,0.175) {};
\node [above] at (0.35,0.175) {$_2$};
\draw node[cross=3pt, very thick] at (0,-0.25) {};
\node [below] at (0,-0.3) {$_3$};
\end{tikzpicture}
&= e^{-{A\over g_s^2}} \, C  \, \Psi^{g=0}_{3\text{-pt}}(\omega_1,\omega_2,\omega_3) \\
&= e^{-{A\over g_s^2}} \, C  \, 2^3 g_s^3 C_{S^2} \, \omega_1 \omega_2 \omega_3,
\fe
where $C_{S^2}$ is a yet undetermined normalization constant associated with the sphere topology and proportional to $g_s^{-2}$.

Likewise, the two next-to-leading order contributions (of order $g_s^2$) to the full four-point cosmological wavefunction that would not involve a delta function from a sphere two-point diagram are from the disconnected product of the torus one-point diagram $\Psi^{g=1}_{1\text{-pt}}(\omega)$ and the sphere three-point diagram $\Psi^{g=0}_{3\text{-pt}}(\omega_1,\omega_2,\omega_3)$,
\ie
\Psi(\omega_1,\omega_2,\omega_3,\omega_4) ~ \supset ~ & 
\exp \Big[\, \begin{tikzpicture}[baseline={([yshift=-.5ex]current bounding box.center)},vertex/.style={anchor=base,
    circle,fill=black!25,minimum size=18pt,inner sep=2pt}]
\draw [thick, fill=black!30] (0,0) circle (0.4);
\draw [thick] (-0.4,0) to [out=-35,in=-145] (0.4,0);
\draw [thick, dashed] (-0.4,0) to [out=15,in=165] (0.4,0);
\end{tikzpicture} \,\Big] 
\exp \Big[\, \begin{tikzpicture}[baseline={([yshift=-.5ex]current bounding box.center)},vertex/.style={anchor=base,
    circle,fill=black!25,minimum size=18pt,inner sep=2pt}]
\draw [thick, fill=black!30] (-0.4,0) to [out=85,in=95] (0.4,0) to [out=-95,in=-85] (-0.4,0);
\filldraw [color=white, fill=white] (-0.15,0.01) to [out=40,in=130] (0.15,0.01) to [out=-160,in=-20] (-0.15,0.01);
\draw [thick] (-0.2,0.05) to [out=-40,in=-130] (0.2,0.05);
\draw [thick] (-0.15,0.01) to [out=40,in=130] (0.15,0.01);
\end{tikzpicture} \,\Big] 
\Big( \begin{tikzpicture}[baseline={([yshift=-.5ex]current bounding box.center)},vertex/.style={anchor=base,
    circle,fill=black!25,minimum size=18pt,inner sep=2pt}]
\draw [thick, fill=black!30] (-0.6,0) to [out=85,in=95] (0.4,0) to [out=-95,in=-85] (-0.6,0);
\filldraw [color=white, fill=white] (-0.15,0.01) to [out=40,in=130] (0.15,0.01) to [out=-160,in=-20] (-0.15,0.01);
\draw [thick] (-0.2,0.05) to [out=-40,in=-130] (0.2,0.05);
\draw [thick] (-0.15,0.01) to [out=40,in=130] (0.15,0.01);
\draw node[cross=3pt, very thick] at (-0.375,0) {};
\node [left] at (-0.5,0) {$_1$};
\end{tikzpicture}
\,\cdot 
\begin{tikzpicture}[baseline={([yshift=-.5ex]current bounding box.center)},vertex/.style={anchor=base,
    circle,fill=black!25,minimum size=18pt,inner sep=2pt}]
\draw [thick, fill=black!30] (0,0) circle (0.4);
\draw [thick] (-0.4,0) to [out=-35,in=-145] (0.4,0);
\draw [thick, dashed] (-0.4,0) to [out=15,in=165] (0.4,0);
\draw node[cross=3pt, very thick] at (-0.2,0.175) {};
\node [above] at (-0.35,0.175) {$_2$};
\draw node[cross=3pt, very thick] at (0.2,0.175) {};
\node [above] at (0.35,0.175) {$_3$};
\draw node[cross=3pt, very thick] at (0,-0.25) {};
\node [below] at (0,-0.3) {$_4$};
\end{tikzpicture}
\,+\, \text{3 permutations \Big)}
\\
& = e^{-{A\over g_s^2}} \, C  \, \frac{g_s^4 C_{S^2} C_{T^2}}{3} \,\omega_1\omega_2\omega_3\omega_4 \left[ 4 + 2 \left( \omega_1^2 + \omega_2^2 + \omega_3^2 +\omega_4^2 \right) \vphantom{\Vhat}\right] ,
\label{eq:full4ptwavef1}
\fe
and from the sphere four-point diagram $\Psi^{g=0}_{4\text{-pt}}(\omega_1,\omega_2,\omega_3,\omega_4)$ 
computed in \cite{Rodriguez:2023kkl}, which contributes through
\ie
\Psi(\omega_1,\omega_2,\omega_3,\omega_4) ~ \supset ~ & 
\exp \Big[\, \begin{tikzpicture}[baseline={([yshift=-.5ex]current bounding box.center)},vertex/.style={anchor=base,
    circle,fill=black!25,minimum size=18pt,inner sep=2pt}]
\draw [thick, fill=black!30] (0,0) circle (0.4);
\draw [thick] (-0.4,0) to [out=-35,in=-145] (0.4,0);
\draw [thick, dashed] (-0.4,0) to [out=15,in=165] (0.4,0);
\end{tikzpicture} \,\Big] 
\exp \Big[\, \begin{tikzpicture}[baseline={([yshift=-.5ex]current bounding box.center)},vertex/.style={anchor=base,
    circle,fill=black!25,minimum size=18pt,inner sep=2pt}]
\draw [thick, fill=black!30] (-0.4,0) to [out=85,in=95] (0.4,0) to [out=-95,in=-85] (-0.4,0);
\filldraw [color=white, fill=white] (-0.15,0.01) to [out=40,in=130] (0.15,0.01) to [out=-160,in=-20] (-0.15,0.01);
\draw [thick] (-0.2,0.05) to [out=-40,in=-130] (0.2,0.05);
\draw [thick] (-0.15,0.01) to [out=40,in=130] (0.15,0.01);
\end{tikzpicture} \,\Big] ~
\begin{tikzpicture}[baseline={([yshift=-.5ex]current bounding box.center)},vertex/.style={anchor=base,
    circle,fill=black!25,minimum size=18pt,inner sep=2pt}]
\draw [thick, fill=black!30] (0,0) circle (0.4);
\draw [thick] (-0.4,0) to [out=-35,in=-145] (0.4,0);
\draw [thick, dashed] (-0.4,0) to [out=15,in=165] (0.4,0);
\draw node[cross=3pt, very thick] at (-0.2,0.175) {};
\node [above] at (-0.35,0.175) {$_2$};
\draw node[cross=3pt, very thick] at (0.2,0.175) {};
\node [above] at (0.35,0.175) {$_3$};
\draw node[cross=3pt, very thick] at (-0.2,-0.2) {};
\node [below] at (-0.35,-0.15) {$_1$};
\draw node[cross=3pt, very thick] at (0.2,-0.2) {};
\node [below] at (0.35,-0.15) {$_4$};
\end{tikzpicture}
~ = ~ e^{-{A\over g_s^2}} \, C  \, \Psi^{g=0}_{4\text{-pt}}(\omega_1,\omega_2,\omega_3,\omega_4)
\\
& = e^{-{A\over g_s^2}} \, C  \, g_s^4 C_{S^2} \, \omega_1 \omega_2 \omega_3 \omega_4 \left[ \alpha + \beta \left( \omega_1^2 + \omega_2^2 + \omega_3^2 + \omega_4^2 \right) \vphantom{\Vhat}\right].
\label{eq:full4ptwavef2}
\fe
where $\alpha\simeq 7.513$ and $\beta\simeq 15.99$.
It is worth noting that the contributions (\ref{eq:full4ptwavef1}) and (\ref{eq:full4ptwavef2}) to the full four-point cosmological wavefunction take precisely the same form.

Lower-point sphere diagrams --- the empty sphere diagram that exponentiates and factors out of every cosmological wavefunction, and the one- and two-point sphere diagrams discussed above --- are rather subtle in the usual formalism of worldsheet string perturbation theory. 
Nevertheless, we hope that existing methods in the literature such as \cite{Tseytlin:1987ww,Erbin:2019uiz,Giribet:2023gub,Anninos:2021ene,Ahmadain:2022tew,Ahmadain:2022eso} can be adapted to the cosmological background (\ref{eq:wscft}) to obtain unambiguous answers for these lower-point diagrams. We leave a more rigorous analysis of these sphere diagrams as well as the empty torus diagram for future research.

Interestingly, the structure of the perturbative expansion of cosmological wavefunctions suggests that the initial wavefunction of the two-dimensional cosmological universe can be reconstructed schematically as\footnote{We thank Xi Yin for this suggestion.} (dropping constant factors for clarity)
\ie
|\Psi_{\text{2D univ.}}\rangle \sim  e^{-{A\over g_s^2}} \, \exp & \bigg[ 
\frac{1}{2}\int \frac{dE}{E} a^\dagger_E a^\dagger_E 
+ g_s \int dE \big(1 + 2E^2 \big) a^\dagger_E 
+ \frac{g_s}{3!}\int dE_1 dE_2 dE_3 a^\dagger_{E_1}  a^\dagger_{E_2} a^\dagger_{E_3} \\
& + \frac{g_s^2}{4!}\int dE_1 dE_2 dE_3 dE_4 \big( \alpha + \beta \sum_{i=1}^4 E_i^2 \big) a^\dagger_{E_1}  a^\dagger_{E_2} a^\dagger_{E_3} a^\dagger_{E_4} +\cdots \bigg] |0\rangle,
\label{eq:wavefuniverse}
\fe
where $|0\rangle$ denotes the perturbative vacuum of strings in the infinite future, and $a^\dagger_E$ denotes a creation operator of a perturbative on-shell string state (\ref{eq:vertexop}) in spacetime with energy $E$, normalized as $[a_E,a_{E'}^\dagger]=E\delta(E-E')$, and the $\cdots$ in (\ref{eq:wavefuniverse}) include corrections that have not yet been computed (for example, the torus two-point diagram at order $g_s^2$). 
We can recognize the leading term in the exponent of (\ref{eq:wavefuniverse}) as that of a squeezed state, as familiar from quantum field theory of free fields in time-dependent backgrounds \cite{Birrell:1982ix,Fulling:1989nb}.\footnote{This is perhaps a more physical argument to expect the two-point sphere diagram to be proportional to an energy-conserving delta function.} String perturbation theory appears to provide a systematic genus expansion in powers of the string coupling $g_s$ on top of this squeezed state. 
It would be interesting to reproduce the structure of the wavefunction (\ref{eq:wavefuniverse}) from the collective field theory of a dual description in terms of a time-dependent solution/background of the $c=1$ matrix quantum mechanics (MQM) \cite{Alexandrov:2002fh,Alexandrov:2003uh,Karczmarek:2003pv}, where $|0\rangle$ corresponds to the vacuum state of the $c=1$ MQM in which one side of the inverted quadratic potential is filled up to the Fermi level $\mu={ 1\over 2\pi g_s}$, and $a^\dagger_E$ corresponds to a creation operator of a collective excitation of the Fermi sea with energy $E$. 
We hope that the computations in this paper help identify the correct MQM dual to the two-dimensional string cosmology, and more broadly to elucidate aspects of quantum field theory and string theory in time-dependent backgrounds.

\section*{Acknowledgements}

The author would like to thank Bruno Balthazar, Minjae Cho, Sergei Dubovsky, Victor Gorbenko, David Kutasov, Savdeep Sethi, and Xi Yin for useful discussions, Scott Collier, Lorenz Eberhardt, and Beatrix Muhlmann for discussions and collaboration on related projects, and Xi Yin for comments on a draft. This research is supported in part by the Simons Collaboration Grant on the Nonperturbative Bootstrap and by the Future Faculty in the Physical Sciences Fellowship at Princeton University.

\appendix

\section{Torus one-point Virasoro conformal blocks}
\label{sec:torusblocks}

In this Appendix we provide the explicit recursion relations that we use to compute the torus one-point Virasoro conformal blocks, as originally derived in \cite{Hadasz:2009db,Cho:2017oxl} and following the notation of \cite{Rodriguez:2023kkl}. 

We parametrize the central charge of the Virasoro algebra as $c=1+6Q^2$ with $Q=b+b^{-1}$, and the holomorphic weights of external primaries as
\ie
h_{\rm ext} = \frac{1}{4}\left( Q^2 - \lambda_{\rm ext}^2 \right).
\label{eq:lami}
\fe
We also define
\ie
h_{r,s} = \frac{1}{4}\left( Q^2 - (rb+sb^{-1})^2 \right), ~~~~~
A_{r,s} = \frac{1}{2}\underset{(p,q)\neq (0,0),(r,s)}{\prod_{p=1-r}^r \prod_{q=1-s}^s } \frac{1}{rb+sb^{-1}}.
\fe
where $q$ is related to the torus modulus $\tau$ by $q=e^{2\pi i \tau}$. The so-called elliptic conformal block $\mathcal{H}_c(h_{\rm ext};h_{\rm int}|q)$ admits a power series expansion in $q$ and obeys the recursion relation,
\ie
\mathcal{H}_{c}(h_{\rm ext};h_{\rm int}|q) = 1 + \sum_{r,s \geq 1} q^{rs} \frac{R_{r,s}}{h_{\rm int}-h_{r,s}} \mathcal{H}_{c}(h_{\rm ext};h_{\rm int}\to h_{r,s} + rs|q),
\label{eq:recrel}
\fe
where the residue $R_{r,s}$ is given by
\ie
R_{r,s}
= A_{r,s} \prod_{k\in\{ 1,2r-1,2 \}} \prod_{l\in\{ 1,2s-1,2 \}} &\frac{\lambda_{\rm ext} + kb + lb^{-1}}{2} \frac{\lambda_{\rm ext} - kb - lb^{-1}}{2} \\ 
&\times\frac{\lambda_{\rm ext} + kb - lb^{-1}}{2} \frac{\lambda_{\rm ext} - kb + lb^{-1}}{2}.
\fe
where the notation $k\in\{ 1,2r-1,2 \}$ stands for $k$ ranging from $1$ to $2r-1$ with step 2.

Using this recursion relation for the torus one-point conformal blocks, one may verify numerically that the one-point functions (\ref{eq:torus1pts}) are modular covariant under an S-transformation of the torus modulus:
\ie
\left\langle V_{P_{\rm ext}}(0) \vphantom{\Vhat}\right\rangle^{T^2(-{1\over \tau})}_{c=25 \text{ Liouv.}} &= |\tau|^{2h_{\rm ext}} \left\langle V_{P_{\rm ext}}(0) \vphantom{\Vhat}\right\rangle^{T^2(\tau)}_{c=25 \text{ Liouv.}}, \\
\left\langle \Vhat_{P_{\rm ext}}(0) \vphantom{\Vhat}\right\rangle^{T^2(-{1\over \tau})}_{c=1 \text{ Liouv.}} &= |\tau|^{2\hhat_{\rm ext}} \left\langle \Vhat_{P_{\rm ext}}(0) \vphantom{\Vhat}\right\rangle^{T^2(\tau)}_{c=1 \text{ Liouv.}},
\label{eq:modcov} 
\fe
where $h_{\rm ext} = 1+P_{\rm ext}^2$ and $\hhat_{\rm ext} = P_{\rm ext}^2$.


\bibliographystyle{JHEP}
\bibliography{2dcosmo1loop1pt_refs}

\end{document}